# How well developed are altmetrics? A cross-disciplinary analysis of the presence of 'alternative metrics' in scientific publications[1]

Zohreh Zahedi[1], Rodrigo Costas[2] and Paul Wouters[3]

[1] *z.zahedi.2@ cwts.leidenuniv.nl,* [2] *rcostas@cwts.leidenuniv.nl,* [3] *p.f.wouters@cwts.leidenuniv.nl,*

Centre For Science and Technology Studies (CWTS), Leiden University,
PO Box 905, 2300 AX, Leiden, The Netherlands

**Abstract**
In this paper an analysis of the presence and possibilities of altmetrics for bibliometric and performance analysis is carried out. Using the web based tool Impact Story, we collected metrics for 20,000 random publications from the Web of Science. We studied both the presence and distribution of altmetrics in the set of publications, across fields, document types and over publication years, as well as the extent to which altmetrics correlate with citation indicators. The main result of the study is that the altmetrics source that provides the most metrics is Mendeley, with metrics on readerships for 62.6% of all the publications studied, other sources only provide marginal information. In terms of relation with citations, a moderate spearman correlation (r=0.49) has been found between Mendeley readership counts and citation indicators. Other possibilities and limitations of these indicators are discussed and future research lines are outlined.

**Introduction**

Citation based metrics and peer review have a long tradition and are widely applied in research evaluation. Citation analysis is a popular and useful measurement approach in the context of science policy and research management. Citations are usually considered as a proxy for 'scientific impact' (Moed 2005). However, citations are not free of limitations (Mac Roberts & Mac Robert 1989, Nicolaisen 2007), they only measure a limited aspect of quality (i.e. the impact on others' scientific publication) (Martin & Irvin 1983; Bornmann & Leydesdorff 2013), their actual meaning has been broadly debated (Wouters 1999) and they also pose technical and conceptual limitations (Seglen 1997; Bordons, Fernandez & Gomez 2002). On the other hand, peer review or peer assessment is also an important instrument and is often regarded as gold standard in assessing the quality of research (Thelwall 2004; Moed 2005; Butler & Macalister 2011;Taylor 2011; Hicks & Melkers 2012), but it has its own limitations and biases as well (Moed 2007; Benos et. al. 2007). Moreover, both citations and peer review are considered mostly as partial indicators of "scientific impact" (Martin & Irvin 1983) and also no single metric can sufficiently reveal the full impact of research (Bollen et. al. 2009). Given these limitations, the combination of peer review with "multi-metric approach" is proposed as necessary for research evaluation (Rousseau & Ye 2013) in the line of the "informed peer review" idea suggested by Nederhof & van Raan (1987).

However, the shortcomings of these more traditional approaches in assessing research have led to the suggestion of new metrics that could inform "new, broader and faster" measures of impact aimed at complementing traditional citation metrics (Priem, Piwowar & Hemminger 2012). This proposal of using and applying so-called 'alternative indicators' in assessing scientific impact has entered the scientific debate, and these new metrics are expected not only to overcome some of the limitations of the previous approaches but also to provide new insights in research evaluation (Priem & Hemminger 2010; Galligan & Dyas-Correia 2013; Bornmann 2013).

These alternative metrics refer to more "unconventional" measures for evaluation of research (Torres-Salinas, Cabezas-Clavijo & Jimenez-Contreras 2013), including metrics such as usage data analysis (download and view counts) (Blecic 1999; Duy & Vaughan 2006; Rowlands & Nicholas 2007; Bollen, Van de Sompel, & Rodriguez 2008; Shuai, Pepe & Bollen 2012); web citation and link analyses (Smith 1999; Thelwall 2001; Vaughan & Shaw 2003; Thelwall 2008; Thelwall 2012) or social web analysis (Haustein 2010). The importance of the web as a rich source for measuring impact of scientific publications and its potentials to cover the inadequacies of current metrics in research evaluation have been also acknowledged in these previous studies. For instance, the scholarly evidence of use of publications found on web are seen as complimentary to citation metrics, also as predictors of later citations (Brody, Harnad & Carr 2006) and being of relevance for fields with less citations

---

(Armbruster 2007). In this sense, the more traditional metrics based on citations, although widely used and applied in research evaluation, are unable to measure the online impact of scientific literature (for example via Facebook, Twitter, reference managers, blogs or wikis) and also lack the ability of measuring the impact of scholarly outputs other than journal articles or conference proceedings, ignoring other outputs such as datasets, software, slides, blog posts, etc. Thus, researchers who publish online and in formats different than journal articles do not really benefit from citation based data metrics.

The rise of these new metrics has been framed with the proposition of the so-called "altmetrics" or social media metrics introduced in 2010 by Priem and colleagues (Priem et. al. 2010) as an alternative way of measuring broader research impacts in social web via different tools (Priem, Piwowar & Hemminger 2012; Priem et. al. 2012). More specifically, altmetrics covers mentions of scientific outputs in social media, news media and reference management tools. This development of the concept of altmetrics has been accompanied by a growth in the diversity of tools that aim to track 'real-time'[2] impact of scientific outputs by exploring the shares, likes, comments, reviews, discussions, bookmarks, saves, tweets and mentions of scientific publications and sources in social media (Wouters & Costas 2012). Among these tools we find F1000 (http://f1000.com), PLOS Article-Level-Metrics (ALM) (http://article-level-metrics.plos.org/), Altmetric.com (www.altmetric.com/), Plum Analytics (www.plumanalytics.com/), Impact Story[3] (www.impactstory.org/), CiteULike (www.citeulike.org/), and Mendeley (www.mendeley.com/). These web based tools capture and track a wide range of researcher's outputs by aggregating altmetrics data across a wide variety of sources. In the next section, we summarize the previous studies on altmetrics that have made use of these tools.

**Background**
The study of altmetrics is in its early stage but some work has already been done. The features of altmetrics tools in general (Zhang 2012) and their validation as a sources of impact assessment has been investigated in some studies. For example, Li & Thelwall & Giustini (2012) studied the strengths, weaknesses and usefulness of two reference management tools for research evaluation. Their findings showed that compared to CiteULike, Mendeley seems to be more promising for future research evaluation. Wouters & Costas (2012) compared features of 16 web based tools and investigated their potentials for impact measurement for real research evaluation purposes. They concluded that although these new tools are promising for research assessment, due to their current limitations and restrictions, they seem to be more useful for self-analysis than for systematic impact measurement at different levels of aggregation.

Shuai, Pepe & Bollen (2012) examined the reactions of scholars to the newly submitted preprints in *arXiv.org*, showing that social media may be an important factor in determining the scientific impact of an article. The analysis of social reference management tools compared to citations has been broadly studied in the field, particularly the comparison of citations and readership counts in Mendeley, in most of the cases showing a moderate and significant correlation between the two metrics (Henning 2010; Priem, Piwowar, & Hemminger 2012; Li & Thelwall & Giustini 2012; Bar-Ilan 2012; Zahedi, Costas & Wouters 2013; Schlögl et. al. 2013; Thelwall et. al. 2013; Haustein et. al. 2013). Also weak correlations between users' tags and bookmarks (as indicators) of journal usage, perception and citations observed for physical journals (Haustein & Siebenlist 2011) have been reported. For the case of F1000, it has been found that both Mendeley user counts and F1000 article factors (FFas) in Genomics and Genetics papers correlate with citations and they are associated with Journal Impact Factors (Li & Thelwall 2012).

Some other studies have focused on whether altmetrics can be used as predictor of citations. For example, in the case of F1000, it has been found that recommendations have a relatively lower predictive power in indicating high citedness as compared to journal citation scores (Waltman & Costas 2013). It has been also suggested that at the paper level, tweets can predict highly cited papers within the first 3 days of publication (Eysenbach 2011) although these results have been criticized by Davis (2012) and more research should delve into this point. Moreover, most of the articles that received blog citations close to their publication time are more highly cited than articles without such blog citations (Shema, Bar-Ilan & Thelwall 2013).

Previous studies mentioned above used altmetrics as a new data source and investigated the association between altmetrics and citation impact. Most of these studies were based on journals such as *Nature & Science (*Li, Thelwall, & Giustini 2012); *JASIST (*Bar-Ilan 2012), Information System Journal (Schlögl et. al. 2013); articles published by bibliometrics and scientometrics community (Bar-Ilan et al. 2012; Haustein, et. al. 2013),

---
[2] Being immediately available compared to citations that take time to accumulate.
[3] Previously known as Total Impact, we use IS in this study to refer to Impact Story. For a review of tools for tracking scientific impact see Wouters & Costas (2012).

*PLoS* and other medical and biomedical journals in PubMed (Priem, Piwowar & Hemminger, 2012; Thelwall et. al. 2013; Haustein, et. al. 2013).

However, to the best of our knowledge, little has been done to date to investigate the presence of altmetrics across various scientific fields and also for relatively ample periods of time. This study is thus one of the first in analyzing a relatively large sample of publications belonging to different fields, document types and publication years. This paper builds upon Wouters & Costas (2012) and Zahedi, Costas & Wouters (2013).

Our main objective in this paper is to present an exploratory analysis of altmetrics data retrieved through Impact Story focusing on the relationship of altmetrics with citations across publications from different fields of science, social sciences and humanities. For this, we examine the extent to which papers have altmetrics obtained through different data sources retrieved via Impact Story and the relationships between altmetrics and citations for these papers. In exploring these issues, we pursue the two following research questions:

1) What is the presence and distributions of Impact Story altmetrics across document types, subject fields and publication years for the studied sample?
2) Is there any relationship between Impact Story-retrieved altmetrics and citation indicators for the studied sample? In other words, to what extent do the Impact Story altmetrics correlate with citation indicators?

**Research methodology**

In this study, we have focused on Impact Story (IS). Although still at an early stage ('beta version'), IS is currently one of the most popular web based tools with some potentials for research assessment purposes (Wouters & Costas 2012). IS aggregates "impact data from many sources and displays it in a single report making it quick and easy to view the impact of a wide range of research output" (http://impactstory.org/faq). It takes as input different types of publication identifiers (e.g. DOIs, URLs PubMed ids, etc.). These are run through different external services to collect the metrics associated with a given 'artifact' (e.g. a publication). A final web based report is created by IS which shows the impact of the 'artifacts' according to a variety of metrics such as the number of readers, bookmarks, tweets, mentions, shares, views, downloads, blog posts and citations in Mendeley, CiteULike, Twitter, Wikipedia, Figshare, Dryad, Scienceseeker, PubMed and Scopus[4].

For this study, we collected a random sample of 20,000 publications with DOIs (published between 2005 and 2011) from all the disciplines covered by the Web of Science (WoS). Publications were randomly collected by using the "*NEW ID ()*" SQL command (Forta 2008, p. 193).The altmetrics data collection was performed during the last week of April 2013. The altmetrics data were gathered automatically via the Impact Story REST API[5], then the responses provided on search requests using DOI's were downloaded. Using this API we could download the altmetric data faster (one request per 18 seconds) compared to the manual data collection we did for the previous study[6]. The files were downloaded per API search request separately in Java Script Object Notations (JSON) format on the basis of individual DOI's and parsed by using the additional JAVA library from within the SAS software[7]. Finally, the data was transformed into a Comma Separated value (CSV) format and matched back with the CWTS in-house version of the Web of Science on the DOIs to be able to add other bibliometric data to them. The final list of publications resulted in 19,772 DOIs (out of 20,000) after matching[8]. Based on this table, we studied the distribution of altmetrics across subject fields, document types and publication years. Citation indicators were calculated and the final files were imported in IBM SPSS Statistics 21 for further statistical analysis.

---

[4] For a full list see http://impactstory.org/faq

[5] A REpresentational State Transfer (REST)(ful) API (Application Programming Interface) used to make a request using GET (DOIs) and collect the required response from impact Story.

[6] In the previous study, the data collection was performed manually directly through the web interface of IS. Manually, IS allowed collecting altmetrics for 100 DOIs per search and maximum 2000 DOIs search per day in order to avoid swamping the limits of its API, for details see Zahedi, Costas & Wouters (2013).

[7] The additional functionality from the "proc groovy" which is a java development environment added to SAS (Statistical analysis Systems) environment for parsing and reading the JSON format and returning the data as an object.

[8] From IS one DOI was missing. We also found that 301 DOIs were wrong in WoS (including extra characters that made them unmatchable, therefore excluded from the analysis). Also 61 original DOIs from WOS pointed to 134 different WOS publications (i.e. being duplicated DOIs). This means that 74 publications were duplicates. Given the fact that there was no systematic way to determine which one was the correct one (i.e. the one that actually received the altmetrics), we included all of them in the analysis with the same altmetrics score resulted in: 20000-1-301+74=19772 final publications. All in all, this process showed that only 1.8% of the initial DOIs randomly selected had some problems, thus indicating that a DOI is a convenient publication identifier although not free of limitations (i.e. errors in DOI data entry, technical errors when resolving DOIs via API and also the existence of multiple publication identifiers in the data sources, resulted in some errors in the full collection of altmetrics for these publications).

In order to test the validity of our sample set we compared the distribution of publications across major fields of science in our sample with that of the whole Web of Science database (Figure 1) in the same period and only those publications with a DOI. As it can be seen, the distribution of publications of our sample basically resembles the distribution of publications in the whole WOS database, so we can consider that our sample is representative of the multidisciplinarity of the database.

**Figure 1. Distribution of publications by major fields of science: sample vs. whole database**

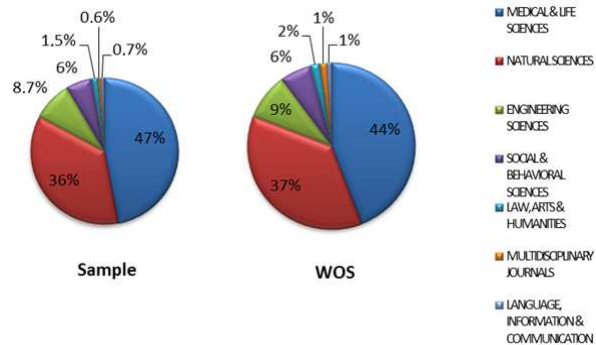

### Results and main findings

In the first place, we present the result of our exploratory analysis of the presence of IS altmetrics over the 19,772 WOS publications published between 2005-2011. Then, we examine the extent to which papers are represented in the data sources both in general and also across document types, subject fields and publication years. Finally, the relationships (correlation) between IS altmetrics and citations for these papers are compared.

### Presence of IS altmetrics by data sources

In our sample, the presence of IS altmetrics across publications is different from each data source. Out of 19,722 publications, 12,380 (62.6%) papers have at least one reader[9] in Mendeley, 324 (1.6%) papers have at least one tweet in Twitter, 289 (1.4%) papers have at least one mention in Wikipedia, 72 (0.3%) papers have at least one bookmark in Delicious and 7413 (37.4%) papers have at least one citation in PubMed. Only 1 paper in the sample has metrics from PLoS ALM[10]. Based on this preliminary test, we decided to exclude some of the metrics from our study: PlosAlm indicators due to their low frequency as they are only available for the PLoS journals thus their presence in our sample is negligible and PubMed-based citations because they are limited only to the Health Sciences and they refer to citations, which we will calculate directly based on the Web of Science. We also decided to sum the metrics coming from Twitter ("Topsy tweets" and "Topsy influential tweets") given their relatively low frequency. As a result, in the current study, the data from Mendeley, Wikipedia, Twitter and Delicious were analyzed.

Table 1 shows the number and percentages of papers with and without IS altmetrics sorted by % of papers with metrics (excluding the PLOS ALM and PubMed metrics). Based on Table 1, our main finding is that, for this sample, the major source for altmetrics is Mendeley, with metrics on readerships for 62.6% of all the publications studied. But for other data sources (Twitter, Wikipedia and Delicious), the presence of metrics across publications is very low, with more than 98% of the papers without metrics. Thus, it is clear that their potential use for the assessment of the impact of scientific publications is still rather limited, particularly when considering a multi-year and multidisciplinary dataset as the one here studied.

**Table 1. Presence of IS altmetrics from data sources**

| Data Source | papers with metrics | % | papers without metrics | % |
|---|---|---|---|---|
| **Mendeley** | 12380 | **62.6** | 7392 | 37.3 |
| **Twitter** | 324 | 1.6 | 19448 | 98.3 |
| **Wikipedia** | 289 | 1.4 | 19483 | 98.6 |
| **Delicious** | 72 | .3 | 19700 | 99.7 |

---

[9] It means that publications without any metrics were left out of the analysis.
[10] This was the only PLOS paper captured by our sample.

**Presence of IS altmetrics across document types**

Regarding document type, out of 19772 publications, there are 16740 (84.7%) articles, 944 (4.7%) review papers, 487 (2.4%) letters and 1601(8%) non-citable[11] items in the sample. Table 2 indicates the coverage of the sampled publications with document types across each data sources. According to Table 2, 81.1% (766) of the review papers, 66.3% (11094) of articles, 25.1% of letters and 24.9% (398) of non-citable in the sample have been saved (read) in the Mendeley. In Twitter, 3.4% (32) of the review papers, 1.9% (30) of non-citable items, 1.5% (255) of articles and 1.4% (7) of letters have tweets. In the case of Wikipedia, 4.6% (43) of the review papers, 1.4% (230) of articles and less than 1% of other document types (letters and non-citable) are mentioned at least once in Wikipedia. Therefore, Mendeley has the highest coverage of all data sources in this sample, (81.1% of the review papers and 66.3% of articles in the sample are covered by Mendeley).

**Table 2. Coverage of publications with different document types by different data sources**

| Doc Type | pub | | Mendeley | | Twitter | | Wikipedia | | Delicious | |
|---|---|---|---|---|---|---|---|---|---|---|
| article | 16740 | 84.7% | 11094 | 66.3% | 255 | 1.5% | 230 | 1.4% | 56 | 0.3% |
| review | 944 | 4.7% | 766 | 81.1% | 32 | 3.4% | 43 | 4.6% | 7 | 0.7% |
| letter | 487 | 2.4% | 122 | 25.1% | 7 | 1.4% | 4 | 0.8% | 3 | 0.6% |
| non-citable | 1601 | 8.0% | 398 | 24.9% | 30 | 1.9% | 12 | 0.7% | 6 | 0.4% |
| **Total** | **19772** | **100** | **12380** | **62.6%** | **324** | **1.6%** | **289** | **1.4%** | **72** | **0.3%** |

We also studied the total numbers of Mendeley readers, tweets, mentions and bookmarks for each document types covered in the sample (i.e. not only the number of publications with metrics, but the frequency of these metrics). Table 3 shows the result of the total sum and the average number of altmetrics scores per document types provided by the different data sources. Based on both table 3 and figure 1, in general, articles have the highest values of numbers of readers, tweets and bookmarks (more than 77.5% of all altmetrics scores are to articles), followed[12] by review papers, non-citables and letters (less than 18% of the altmetrics scores are to the other types) in all data sources. But considering the average metrics per publications[13], it can be seen that, Mendeley accumulate the most metrics per all document types than all other data sources. Also, in Mendeley, review papers have attracted the most readers per publications (on average there are ~14 readers per review paper) than all other data sources.

**Table 3. Distribution of IS altmetrics per document types in different data sources**

| Doc Type | pub | Mendeley Readers | % | Avg | Tweets | % | Avg | Wikipedia Mentions | % | Avg | Delicious Bookmarks | % | Avg |
|---|---|---|---|---|---|---|---|---|---|---|---|---|---|
| Article | 16740 | 82553 | 83.3 | 4.9 | 3020 | 94.5 | 0.18 | 292 | 77.5 | 0.02 | 213 | 87.3 | 0.01 |
| Review | 944 | 12730 | 12.9 | 13.4 | 78 | 2.4 | 0.08 | 68 | 18.0 | 0.07 | 7 | 2.9 | 0.01 |
| Non-citable | 487 | 3301 | 3.3 | 2.0 | 76 | 2.4 | 0.05 | 13 | 3.4 | 0.01 | 14 | 5.7 | 0.01 |
| Letter | 1601 | 466 | 0.5 | 0.9 | 21 | 0.7 | 0.04 | 4 | 1.1 | 0.01 | 10 | 4.1 | 0.02 |
| **Total** | **19772** | **99050** | **100** | **5.0** | **3195** | **100** | **0.16** | **377** | **100** | **0.02** | **244** | **100** | **0.01** |

---

[11] Non-citable document type corresponds to all WOS document types other than article, letter and review (e.g. book reviews, editorial materials, etc.).
[12] in Delicious, articles, non-citables, letters and review papers have the highest number of metrics orderly.
[13] Average metrics per publications calculated by dividing the total numbers of metrics from each data source by total number of publications in the sample. For example, in Mendeley, average number of readers per publication equals to 99050/19772=~5

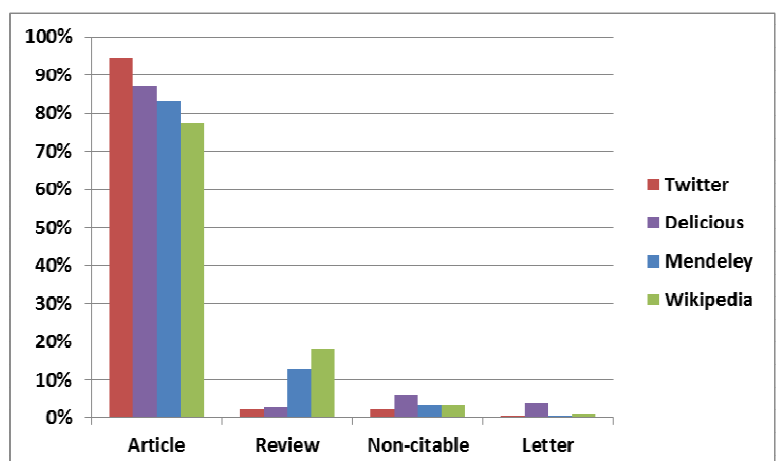

**Figure 1. Distribution of IS altmetrics across document types**

**Presence of IS altmetrics across NOWT Subject fields**
For this analysis, we used the NOWT (High) classification which has 7 major disciplines developed by CWTS[14]. Table 4 shows the percentage of publications having at least one metrics (i.e. papers with at least one reader in Mendeley, once bookmarked in Delicious, once tweeted, or once mentioned in Wikipedia) across those major disciplines[15]. According to the results, Multidisciplinary publications ranked the highest in all data sources. The major source for altmetrics data in our sample is Mendeley with the highest proportion for Multidisciplinary fields, which include journals such as Nature, Science or PNAS. 80% of the publications in this field, 73% of the publications from Medical & Life Sciences[16] and 68% of the publications from Social & Behavioural Sciences have at least one Mendeley reader. Among the other data sources, Multidisciplinary publications ranked the highest as well but with lower presence of publications with metrics. Regarding the top three fields with the highest percentage of altmetrics, Wikipedia has similar pattern as Mendeley: 7% of the publications from Multidisciplinary field, 2% of the publications from Medical & Life Sciences and 2% of the publications from Social & Behavioural Sciences have at least one mention in Wikipedia. In Twitter, 7% of the publications from Multidisciplinary field, 3% of the publications from Social & Behavioural Sciences and 2% of publications from Medical & Life Sciences are the top three fields that have at least one tweet. In Delicious, only 1% of the publications from Multidisciplinary field, Language, Information & Communication and Social & Behavioural Sciences have at least one bookmark while other fields have less than 1% altmetrics.

**Table 4. Coverage of publications with different NOWT subject fields by different data sources**

| NOWT High Subject Categories | Total number of publications | | Mendeley | | Wikipedia | | Twitter | | Delicious | |
|---|---|---|---|---|---|---|---|---|---|---|
| MULTIDISCIPLINARY JOURNALS | 216 | 47% | 172 | 80% | 15 | 7% | 16 | 7% | 3 | 1% |
| MEDICAL & LIFE SCIENCES | 15637 | 36% | 11353 | 73% | 284 | 2% | 301 | 2% | 67 | 0.4% |
| SOCIAL & BEHAVIORAL SCIENCES | 1878 | 6% | 1268 | 68% | 32 | 2% | 58 | 3% | 11 | 1% |
| NATURAL SCIENCES | 11935 | 8.7% | 6554 | 55% | 103 | 1% | 123 | 1% | 34 | 0.3% |
| ENGINEERING SCIENCES | 2885 | 0.6% | 1558 | 54% | 7 | 0.2% | 9 | 0.3% | 2 | 0.1% |
| LANGUAGE. INFORMATION & COMMUNICATION | 241 | 0.7% | 123 | 51% | 2 | 1% | 1 | 0.4% | 3 | 1% |
| LAW. ARTS & HUMANITIES | 488 | 1.5% | 190 | 39% | 8 | 2% | 7 | 1% | 0 | 0% |
| | | 100 | | | | | | | | |

---

[14] In the previous study, we used the NOWT (Medium) with 14 subject fileds. For more details see: http://nowt.merit.unu.edu/docs/NOWT-WTI_2010.pdf
[15] Here publications can belong to multiple subject categories.
[16] According to the Global Research Report by Mendeley (http://www.mendeley.com/global-research-report/#.UjwfTsanqgk), coverage of Mendeley in different subjects are as follows: the highest coverage are by publications from Biological Science & Medicine (31%), followed by Physical Sciences and Maths (16%), Engineering & Materials Science (13%), Computer & Information Science (10%), Psychology, Linguistics & Education(10%), Business Administration, Economics & Operation Research (8%), Law & Other Social Sciences (7%) and Philosophy, Arts & Literature & other Humanities (5%)

Again, the total scores of Mendeley readers, tweets, mentions and bookmarks for each discipline in the sample have been calculated. Figure 2 shows that the distributions of IS altmetrics across different subject fields is uneven. Both Medical & Life and Natural Sciences received the highest proportion of altmetrics in all data sources. In general in all data sources, more than 30% of altmetrics accumulated by publications from Medical & Life Sciences and more than 23% of altmetrics are to publications from the fields of Natural Sciences. Other fields, each received less than 10% of total altmetrics. Comparing the different data sources in terms of the proportion of altmetrics across fields, different patterns arise: Medical & Life Sciences fields proportionally attracted the most attention in Wikipedia, followed by Mendeley, Twitter and Delicious while in case of Natural Sciences, Delicious, Twitter, Mendeley and Wikipedia, proportionally got the most attention orderly; moreover, for Mendeley, both Social & Behavioural and Engineering Sciences, proportionally, received the highest attention than all other fields.

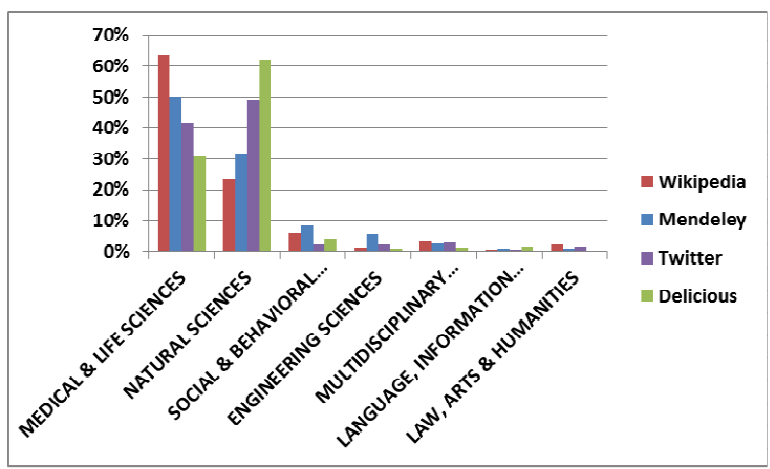

**Figure 2. Distribution of IS altmetrics across NOWT subject fields**

**Table 5. Distribution of IS altmetrics per NOWT subject fields in different data sources**

| NOWT Subject category | Mendeley Readers | | Wikipedia Mentions | | Delicious Bookmarks | | Tweets | |
|---|---|---|---|---|---|---|---|---|
| **MEDICAL & LIFE SCIENCES** | 86347 | 50% | 371 | 64% | 92 | 31% | 1958 | 42% |
| **NATURAL SCIENCES** | 54481 | 32% | 136 | 23% | 186 | 62% | 2317 | 49% |
| **SOCIAL & BEHAVIORAL SCIENCES** | 14102 | 8% | 35 | 6% | 12 | 4% | 112 | 2% |
| **ENGINEERING SCIENCES** | 9800 | 6% | 7 | 1% | 2 | 1% | 100 | 2% |
| **MULTIDISCIPLINARY JOURNALS** | 4521 | 3% | 20 | 3% | 3 | 1% | 144 | 3% |
| **LANGUAGE, INFORMATION & COMMUNICATION** | 1492 | 0.9% | 2 | 0.3% | 4 | 1% | 1 | 0% |
| **LAW, ARTS & HUMANITIES** | 1297 | 0.8% | 13 | 2% | 0 | 0% | 72 | 2% |
| | | 100 | | 100 | | 100 | | 100 |

**Comparison of Citations per Papers (CPP) and Readerships per Papers (RPP) across fields**
Although measuring the impact of scholarly publications in social media is very important, it is not yet clear for what purposes scholarly publications are mentioned in social media and reference management tools such as Mendeley, in social bookmark manager such as Delicious, in Wikipedia and Twitter by different users/scholars, and particularly it is not clear if these mentions can be considered as measures of any type of "impact" of the publications. In case of Mendeley, it is assumed that publications are saved in users' libraries for immediate or later reading and possibly also future citation.

In any case, it is important to know how many altmetrics vs. citations each publication received and what are the different pattern across different subject fields. Due to the fact that not all of scholarly publications are covered equally by citation databases and also the existence of disciplinary differences in terms of citations, which vary a lot between fields, it is interesting to study both the proportion of altmetrics vs citations per publications to see which fields can benefit from having more density of altmetrics scores (i.e. altmetrics scores per paper) than citation density. Since Twitter, Wikipedia and Delicious showed an overall very low presence per paper, we focus here only on Mendeley. Both the average number of Mendeley readerships per papers (RPP) and WOS citations per papers (CPP) across different NOWT subject fields were calculated and analyzed (Figure 3). For calculating the citations (excluding self-citations), we used a variable citation window from the year of publication to 2012. Also a variable "readership window" was considered for Mendeley, counting readerships from the publication year of the paper until the last week of April 2013. In this analysis we have also included publications without any metrics (citations or Mendeley readers). The result (Figure 3 sorted by RPP) shows that in general, Multidisciplinary journals have the highest values of both RPP and CPP; and Law, Arts & Humanities have the lowest values. For fields such as Multidisciplinary journals, Medical & Life Sciences, Natural and Engineering Sciences, the value of CPP is higher than RPP, while for fields such as Social & Behavioural Sciences, Language, Information & Communication and Law, Arts & Humanities, RPP outperforms CPP. The latter is an interesting result that might suggest the relevance of Mendeley for the study of Social Sciences and Humanities publications, which are often not very well represented by citations (Nederhof 2006). In order to further test the differences between RPP and CPP, we extended the same type of analysis for all 248 WOS individual subject categories, resulting that 167 out of 248 WOS subject categories have higher CPP values than RPP values. Most of the fields with higher values of CPP vs. RPP are from the Sciences (145), 18 from the Social Sciences and 4 from the Art and Humanities. On the other hand, 72 fields presented higher RPP than CPP scores (among them 31 are from Social Sciences, 27 from Science and 13 from Art and Humanities) [17]. Therefore, we can conclude that citations are more dominant than readerships particularly in the fields of the Sciences (which are also the fields with the highest coverage in citation databases); while on the other hand, many sub-fields from the Social Sciences and Art and humanities received proportionally more readerships per paper than citations per paper. This could be seen as a possibility for these fields with lower coverage in citation databases (such as WoS) to benefit from Mendeley in terms of having more readership impact than citation impact, although this needs further explorations.

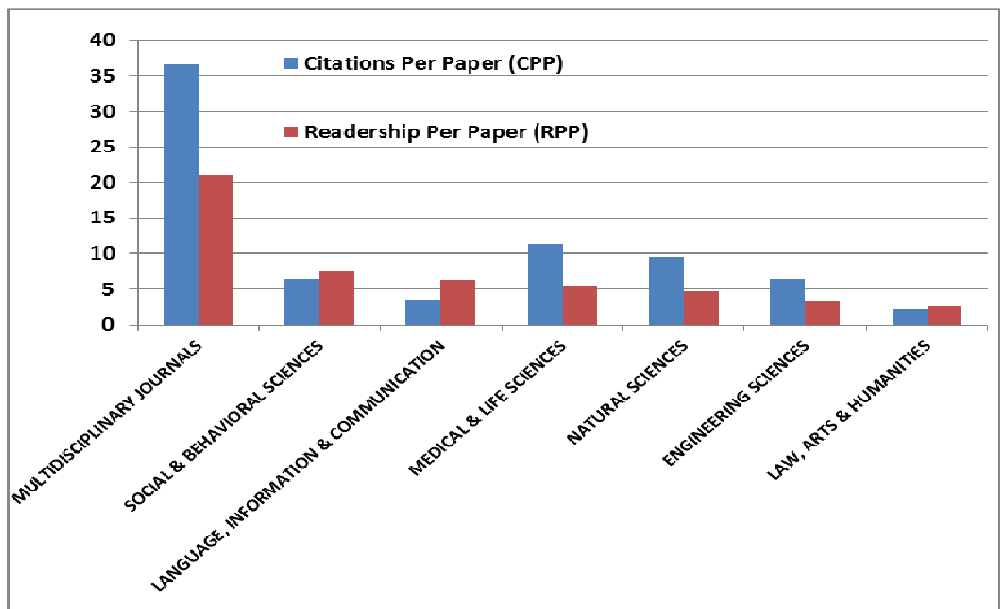

**Figure 3. Comparing CPP and RPP in Mendeley across Subject Fields**

**Trend analysis of IS altmetrics across publication years**

Table 6 shows the trend analysis of number and share of publications in the sample by altmetrics sources. Regarding the publication years, the share of publications ranges from 10% in the year 2005 to 18% in the year 2011. The coverage of different sources is also shown in the table. In our sample, Mendeley has its peak in its proportion of publications with some readers in 2009 (66%) and the lowest point in 2011 (57%), although the total number of publications with some Mendeley readerships has increased during the whole period, with the exception of 2011 when there is a small drop compared to 2010. Twitter has its highest peak in 2011 (4%) and its lowest values in the early years (around 1% between 2005-2009). Wikipedia mentions are for 2% of all the publications published between 2005 to 2008 and 1% of all the publications published between 2010 and 2011. For Delicious, the highest peak is for the years 2007 and 2011 and the lowest one for the year 2005, also publications from 2008, 2009 and 2010 have the same presence in Delicious. All in all, it seems that Twitter and Delicious tend to cover the more recent publications better than the older ones although the values are in general very low.

---

[17] For 9 fields (8 fields from Art and Humanities and 1 field from Science) CPP and RPP scores were exactly the same

**Table 6. Coverage of publications with different publication years by different data sources**

| Pub year | p | | Mendeley | | Wikipedia | | Delicious | | Twitter | |
|---|---|---|---|---|---|---|---|---|---|---|
| **2005** | 2006 | 10% | 1263 | 63% | 39 | 2% | 3 | 0.1% | 17 | 1% |
| **2006** | 2405 | 12% | 1491 | 62% | 58 | 2% | 4 | 0.2% | 6 | 0.2% |
| **2007** | 2682 | 14% | 1702 | 63% | 41 | 2% | 13 | 0.5% | 16 | 1% |
| **2008** | 2858 | 14% | 1799 | 63% | 46 | 2% | 11 | 0.4% | 34 | 1% |
| **2009** | 3039 | 15% | 2001 | 66% | 43 | 1% | 12 | 0.4% | 31 | 1% |
| **2010** | 3228 | 16% | 2099 | 65% | 37 | 1% | 13 | 0.4% | 62 | 2% |
| **2011** | 3548 | 18% | 2020 | 57% | 25 | 1% | 16 | 0.5% | 158 | 4% |

The presence of overall altmetrics scores (i.e. not only publications with altmetrics, but their total counting) has been also calculated in order to know its trend over time. According to Table 7, this is quite different across different data sources. For example, for Wikipedia and Mendeley, publications from the years 2006 and 2009, accumulated most of the mentions (20%) and readerships (17%) respectively. In the case of Mendeley and Wikipedia we noticed a decrease in the amount of altmetrics in the last two years.

Both in Delicious and in Twitter, publications from the year 2008 received the highest proportion of altmetrics. In case of Delicious, 50% of bookmarks and in case of Twitter, 34% of tweets are to publications published in 2008. Comparing the amount of altmetrics in each year across different data sources shows that in this sample, both the oldest and the most recent publications in Twitter have the most altmetrics (tweets) (26% of tweets are to publications from the year 2005[18] and 2011 respectively) and also the recent publications (2009-2010) have the most altmetrics (readerships) in Mendeley (figure 4).

**Table 7. Distribution of IS Altmetrics across publication year**

| Pub year | p | | Mendeley | | Wikipedia | | Delicious | | Twitter | |
|---|---|---|---|---|---|---|---|---|---|---|
| **2005** | 2006 | 10% | 10814 | 11% | 48 | 13% | 51 | 21% | 835 | 26% |
| **2006** | 2405 | 12% | 12658 | 13% | 77 | 20% | 4 | 2% | 20 | 1% |
| **2007** | 2682 | 14% | 13739 | 14% | 58 | 15% | 14 | 6% | 102 | 3% |
| **2008** | 2858 | 14% | 14299 | 14% | 67 | 18% | 122 | 50% | 1072 | 34% |
| **2009** | 3039 | 15% | 16922 | 17% | 50 | 13% | 21 | 9% | 145 | 5% |
| **2010** | 3228 | 16% | 16305 | 16% | 43 | 11% | 14 | 6% | 198 | 6% |
| **2011** | 3548 | 18% | 14239 | 14% | 34 | 9% | 18 | 7% | 823 | 26% |
| | | 100% | | 100% | | 100% | | 100% | | 100% |

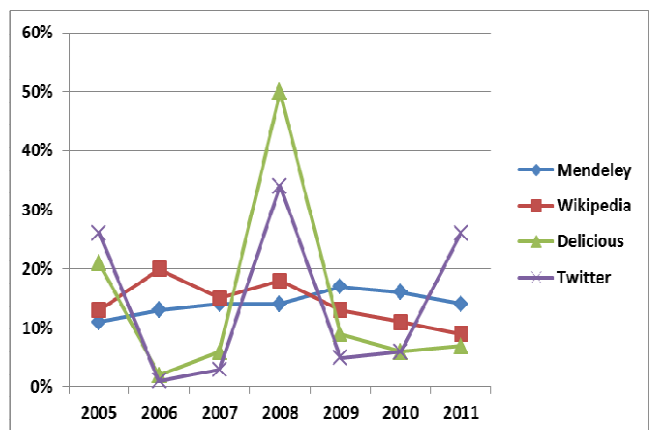

**Figure 4. Distribution of IS altmetrics across publication years**

---

[18] In 2005, the two most tweeted papers are from the field of Physics, they received more than half of the total tweets in this year (472 tweets), thus showing a strong skewed distribution.

**Relationships between IS altmetrics and citation indicators**
In this section we study more thoroughly the relationship between the IS altmetrics and citation indicators. Following the CWTS standard calculation of indicators (cf. Waltman et. al. 2011), we calculated for all the publications the following citation indicators: *Citation Score* (CS), that is, number of citations per publications; *Normalized Citation Score* (NCS), that is, number of citations per publications, with a normalization for fields differences and publication year; *Journal Citation Score* (JCS), that is the average number of citations received by all publications in that journal of a publication; and *Normalized Journal Score* (NJS), that is, the average number of citations received by all publications in that journal normalized by fields differences and publication year. For the calculation of the impact indicators, as explained before, we used a variable citation window (i.e. citations up to 2012) excluding self-citations. The result of the factor analysis, the correlation analysis and impact of publications with and without altmetrics will be presented in the next sections.

**Factor analysis of IS altmetrics and bibliometrics indicators**
An exploratory factor analysis has been performed using SPSS version 21 in order to know more about the underlying structure, relationship among the variables and the dimension of variables (Table 8). Principal Component Analysis (PCA) revealed the presence of 2 main components or dimensions with eigenvalues exceeding 1, explaining 58% of the total variance. The first dimension is dominated by bibliometric indicators. Mendeley readerships and Wikipedia mentions are also included in this dimension; although Mendeley readership counts has the highest loadings in this dimension of the two indicators. The second dimension is more related to social media metrics, showing that Twitter and Delicious are strongly correlated. These results suggest that the variables in each group may represent similar concepts.

**Table 8. Factor analysis of the variables**

| Rotated Component Matrix a | | |
|---|---|---|
| | Component | |
| | 1 | 2 |
| **CS** | **.837** | *.005* |
| **NCS** | **.752** | *.009* |
| **JS** | **.745** | *-.011* |
| **NJS** | **.720** | *-.015* |
| **Mendeley** | **.680** | *.008* |
| **Wikipedia** | **.297** | *.009* |
| **Delicious** | *.003* | **.954** |
| **Twitter** | *.004* | **.954** |
| Extraction Method: Principal Component Analysis. Rotation Method: Varimax with Kaiser Normalization. Loadings higher than .1 are shown. 58% of total variance explained. | | |
| a. Rotation converged in 4 iterations. | | |

**Correlations between IS altmetrics and bibliometrics indicators**
In order to overcome the technical limitation of SPSS for calculating Spearman correlation for large datasets[19], first, rankings of variables computed using Data>rank cases and then Pearson correlation performed on the ranked variables; this method provides the spearman correlation of the original variables. Table 9 shows the result of the correlation analysis among the different altmetrics data source and citation and journal citation scores and their 95% confidence intervals (calculated using the Bootstrapping technique implemented in SPSS). According to this table, citation indicators are more correlated between them than with altmetrics. In general, direct citations indicators (i.e. CS and NCS) correlate better among them than with indicators of journal impact (JS and NJS), although the correlations between the two groups are fairly high. Mendeley is correlated with Wikipedia (r=.08) and Twitter is correlated with Delicious (r=.12), this is in line with the result of the factor analysis but the correlation values are very low. Compared to citation indicators, Mendeley has the highest correlation score with citations (moderate correlation of r=0.49) among all the altmetrics sources. The other altmetric sources show very weak or negligible correlation with citation indicators.

---

19 Calculating Spearman correlation analysis in SPSS for large datasets gives this error: "Too many cases for the available storage", for overcoming this limitation, we followed the process we mentioned in the text. For more details see: http://www-01.ibm.com/support/docview.wss?uid=swg21476714

**Table 9. Correlation analysis of the rank values of variables**

|  | NCS | JS | NJS | Mendeley | Wikipedia | Delicious | Twitter |
|---|---|---|---|---|---|---|---|
| **CS** | .886 (.882-.89) | .762 (.756-.769) | .557 (.547-.567) | .497 (.485-.508) | .094 (.08-.108) | .011 (-.005-.027) | .025 (.01-.039) |
| **NCS** |  | .528 (.516-.538) | .6 (.59-.609) | .467 (.455-.478) | .074 (.059-.087) | .019 (.002-.035) | .054 (.037-.068) |
| **JS** |  |  | .711 (.702-.718) | .44 (.428-.452) | .09 (.075-.105) | -.003 (-.018-.012) | -.003 (-.018-.011) |
| **NJS** |  |  |  | .427 (.415-.439) | .058 (.044-.072) | .012 (-.005-.028) | .039 (.023-.053) |
| **Mendeley** |  |  |  |  | .083 (.067-.099) | .031 (.015-.047) | .07 (.055-.084) |
| **Wikipedia** |  |  |  |  |  | .021 (-.001-.049) | .056 (.025-.087) |
| **Delicious** |  |  |  |  |  |  | .125 (.073-.185) |

**Impact of publications with/without altmetrics**

In this section, we study the differences in impact between publications with and without altmetrics. The main idea is to see whether publications with altmetrics tend to have more citation impact than those without altmetrics. Table 10 presents the bibliometric indicators and their 95% confidence intervals (calculated using the Bootstrapping technique implemented in SPSS). For instance, according to the median values it can be observed that publications with metrics have in general higher citation scores compared to those without metrics in all data sources (although, in some cases, the confidence intervals show some overlapping, thus the claim of the higher impact for these cases is less strong and probably more influenced by outliers).

**Table 10. Comparison of NCS and NJS of the publications with and without altmetrics**

|  |  | With Metrics | | | | Without Metrics | | | |
|---|---|---|---|---|---|---|---|---|---|
|  |  | CS | JS | NCS | NJS | CS | JS | NCS | NJS |
| **Mendeley** | N | 12380 | 12380 | 12380 | 12380 | 7392 | 7392 | 7392 | 7392 |
|  | Median | 5 | 6.53 | 0.72 | 1.02 | 1 | 1.76 | 0.10 | 0.53 |
| Confidence Interval | Lower | 4 | 6.4 | 0.69 | 1.01 | 0.5 | 1.67 | 0.08 | 0.51 |
|  | Upper | 5 | 6.69 | 0.74 | 1.04 | 1 | 1.89 | 0.12 | 0.55 |
| **Wikipedia** | N | 289 | 289 | 289 | 289 | 19483 | 19483 | 19483 | 19483 |
|  | Median | 12 | 13.87 | 1.18 | 1.18 | 2 | 4.43 | 0.47 | 0.86 |
| Confidence Interval | Lower | 9 | 11.91 | 0.97 | 1.07 | 2 | 4.32 | 0.46 | 0.85 |
|  | Upper | 14 | 15.2 | 1.35 | 1.31 | 3 | 4.57 | 0.49 | 0.87 |
| **Twitter** | N | 324 | 324 | 324 | 324 | 19448 | 19448 | 19448 | 19448 |
|  | Median | 4 | 3.6 | 1 | 1.1 | 3 | 4.53 | 0.47 | 0.86 |
| Confidence Interval | Lower | 3 | 3.1 | 0.86 | 0.97 | 2 | 4.39 | 0.46 | 0.85 |
|  | Upper | 5 | 4.74 | 1.29 | 1.27 | 3 | 4.62 | 0.49 | 0.87 |
| **Delicious** | N | 72 | 72 | 72 | 72 | 19700 | 19700 | 19700 | 19700 |
|  | Median | 3 | 3.99 | 0.89 | 1.07 | 3 | 4.52 | 0.48 | 0.86 |
| Confidence Interval | Lower | 2 | 2.34 | 0.52 | 0.76 | 2 | 4.38 | 0.46 | 0.85 |
|  | Upper | 6 | 5.55 | 1.57 | 1.33 | 3 | 4.62 | 0.49 | 0.87 |

Focusing on the number of Mendeley readers per publication and considering their impact as measured by the NCS and NJS, we can see how publications tend to increase in citation impact as the number of readerships increases (Figure 5). The effect is quite strong, especially for the average number of citations per publication but this is less prominent for the NJS indicator. The same result found by Waltman & Costas (2013) for relationship between recommendations from F1000, citations and journal impact. In their study, they found that on average, publications with more recommendations also have higher citation and journal impact.

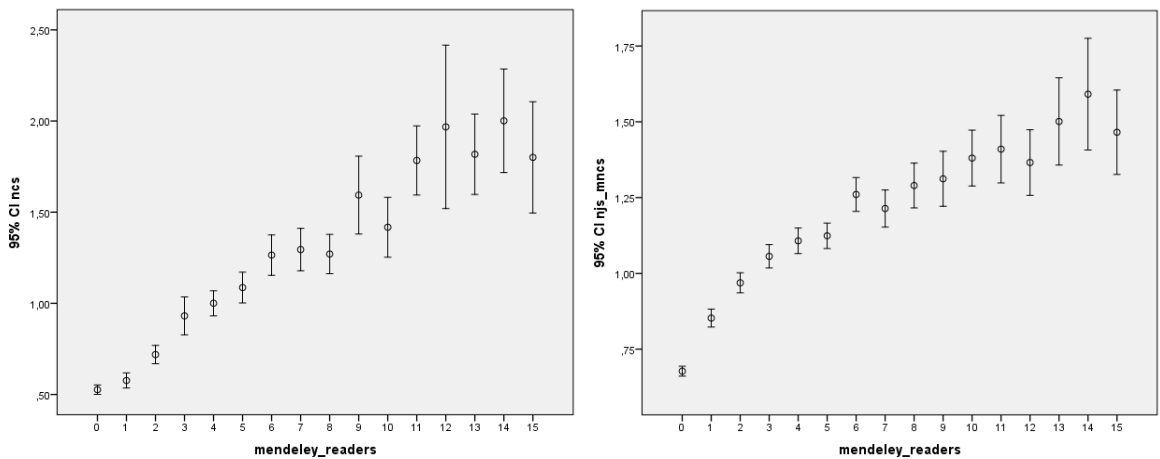

**Figure 5. Relation between number of Mendeley readerships and citation and journal impact**

**Discussion and conclusions**
In this paper we have used Impact Story[20] for gathering altmetrics for a set of randomly sampled publications. IS is an interesting open source for collecting altmetrics, however, we also see some important limitations[21] particularly regarding the speed and capacity of data collection and formatting of the data. We detect different results comparing our current results with those presented in our previous study (Zahedi, Costas & Wouters 2013) mostly due to the different methodology of data collection (manually vs. automatically) and collecting the data at different points in time as it happened between our two studies, where in the first one, Mendeley was only presented in around 37% of the publications[22] and now in more than 60%[23]. This situation also points to the need for the tools to be transparent in how their data are collected and their limitations. This means that an important natural future step will be the proper assessment of the validity of the data retrieved via different altmetrics data sources (as it has been done for example for Google Scholar – cf. Delgado López-Cózar et. al. 2012). This validation of the quality, reliability and robustness of the altmetrics tools is essential in order to be able to apply altmetrics for serious research assessment purposes. For these tools to be fully incorporated in regular research assessment processes, they need to meet the necessary requirements for data quality, transparency and indicator reliability and validity as emphasized by Wouters & Costas (2012) in their study of altmetric tools. Moreover, the results of this study are based on the WOS covered publications; hence, it is important to keep in mind the restrictions of this database with regards to its coverage of some fields, language and publication formats (Moed 2009; Van Raan, Van Leeuwen & Visser 2011; Archambault & Larivière 2006; Torres-Salinas, Cabezas-Clavijo & Jimenez-Contreras 2013).

All in all, given the exploratory nature and the fact that basically the same results have been found with the two data collections, we can assume that our results are robust and valid for our purposes. In general, our study shows that Mendeley is the major and more useful source for altmetrics data. Mendeley has the highest coverage and proportion of altmetrics compared to Twitter, Wikipedia and Delicious for the studied publications. Out of 19,772 publications a total 12380 cases (62.6%) had at least one reader in Mendeley. Previous studies also showed that Mendeley is the most exhaustive altmetrics data source (Bar-Ilan et. al. 2012, Priem et. al. 2012) mostly for the publications from Library and Information Science field: 97.2% coverage for JASIST articles published between 2001 and 2011 (Bar-Ilan 2012); 82% coverage for articles published by researchers in Scientometrics (Bar-Ilan et al. 2012); and 82% of bibliometrics literature (Haustein et. al. 2013), for Multidisciplinary journals such as Nature and Science (94% and 93% of articles published these journals in 2007) (Li, Thelwall and Giustini 2012); and more than 80% of PLoS ONE publications (Priem et. al. 2012) covered by Mendeley. In terms of document type, review papers and articles were proportionally the most read, shared, liked or bookmarked format compared to non-citable items and letters across all data sources. Multidisciplinary fields (i.e. the field where journals such as *Nature*, *Science* or the *PNAS* are included) are the most present in all altmetrics data sources but concerning the distribution of altmetrics across different fields, more than 30% of altmetrics accumulated by publications from Medical & Life Sciences and more than 23% of altmetrics are to publications from the fields of Natural Sciences. Comparing both proportion and distribution of IS altmetrics across different fields among different data sources shows different patterns, particularly in Mendeley, both Social & Behavioural and Engineering Sciences, have proportionally received the highest attention compared to all other fields. Considering citations and readerships per publication, Multidisciplinary journals have the highest and Law, Arts & Humanities have the lowest density of both citations and readerships per publications. However, according to our observation, there is a higher density of readerships per paper than citations per papers in several fields of the Social Sciences and Humanities. This finding suggests that Mendeley readership counts could have some added value in supporting the evaluation and analysis of these fields, which have been traditionally worse represented by citation indicators (cf. Nederhof 2006). Another explanation for those fields with lower proportion of readers than citations could be the fact that Mendeley is relatively new and not yet widely used and adopted among all scholars from all the disciplines. Besides, differences in citation and readership behaviors and practices among fields could also explain these differences. In any case, this is an aspect that needs further analysis.

---

20 Impact Story, was in an initial stage of development (i.e. in a 'Beta' version) at the moment of development of this study.

21 For current limitations of IS see: http://impactstory.org/faq#toc_3_11

22 The time interval between the first and the second data collection was 6 months and data collection done manually versus the second one which done automatically using RESTAPI calls.

23 Reasons for these differences can be the changes/improvements in the identification of publications by Mendeley (e.g. by merging version of the same paper, identifying more DOIs, increments in the number of users in Mendeley, etc.

Our trend analysis shows that particularly publications with Mendeley readerships have increased over time, although there is a slight decrease in the number of readerships and proportion of publications with Mendeley readers for the last two years. The most plausible explanation for this is that the accumulation of readers takes some time. To the best of our knowledge there is no information on the 'readership history' of publications (besides the fact that readerships could conceptually decrease as the users delete or change their libraries) and so far we don't have results on the readerships pace. This means that we don't know when a paper in a given year has obtained its peak in readerships. It is highly likely, that although faster than citations, the accumulation of readerships for publications also takes some time, and this is the reason why for the most recent publications, the number of readers is slower as compared to those older publications that have had more time to accumulate readerships. Future research should also focus on disentangling this aspect.

The Spearman correlation of Mendeley readerships with citation impact indicators showed moderate correlations (r=.49) between the two variables which is also found in other previous studies (Bar-Ilan 2012; Priem et. al. 2012). This indicates that reading and citing are related activities, although still different activities that would be worthwhile to explore. According to the result of comparing the impact of publications with and without altmetrics with their citation scores, it can be also concluded that in general, publications with more altmetrics also tend to have both higher direct citations and are published in journals of higher impact. The issue about the potential predictability of citations through altmetric scores will be explored in follow-up research.

Finally, although citations and altmetrics (particularly Mendeley readerships) exhibit a moderate positive relationship, it is not yet clear what the quality of the altmetrics data is and neither what kind of dimension of impact they could represent. Since altmetrics is still in its infancy, at the moment, we don't yet have a clear definition of the possible meanings of altmetric scores. In other words, the key question of what altmetrics mean is still unanswered. From this perspective, it is also necessary to know the motivations behind using these data sources, for example in case of Mendeley: what does it reflect when an item is saved/added by several users to their libraries? Also, what does it mean that an item is mentioned in Wikipedia, CiteULike, Twitter and any other social media platform? Does it refer to the same or different dimension compared to citation? In the same line, besides studying to what extent different publications are presented in Mendeley and other social media tools and their relations with citation impact, we need to study for what purposes and why these platforms are exactly used by different scholars. Moreover, research about the quality and reliability of the altmetric data retrieved by the different altmetrics providers is still necessary before any interpretation and potential real uses for these data and indicators are developed. This information in combination with the assessment of the validity and reliability of altmetrics data and tools will shed more light on the meanings of altmetrics and can help to unravel the hidden dimensions of altmetrics in future studies.

**Acknowledgement:**
This study is the extended version of our research in progress paper (RIP) presented at the 14[th] International Society of Scientometrics & Informetrics Conference (ISSI) Conference, 15-19 July, 2013, Vienna, Austria. We thank the Impact Story team for their support in working with the Impact Story API. This work is partially supported by the EU FP7 ACUMEN project (Grant agreement: 266632). The authors would like to thank Erik

Van Wijk from CWTS for his great help in managing altmetrics data. The authors also acknowledge the useful suggestions of Ludo Waltman from CWTS and the fruitful comments of the anonymous referees of the journal.